\documentclass{aa}  
\usepackage{graphicx}
\usepackage{txfonts}
\usepackage{natbib}
\newcommand {\HII}    {H\,{\sc ii} }

      \def\new#1 {{\bf #1 }}
      \def\cut#1 {\sout{#1} }
\begin{document}
%
   \title{Observations of the Goldreich-Kylafis effect in star-forming regions with XPOL at the IRAM 30m telescope}
   \titlerunning{Observations of the Goldreich-Kylafis effect}

   \author{J. Forbrich\inst{1,2} \and H. Wiesemeyer\inst{3} \and C. Thum\inst{4} \and A. Belloche\inst{1} \and K.~M. Menten\inst{1}}
   \institute{Max-Planck-Institut f\"ur Radioastronomie, Auf dem H\"ugel 69, D-53121 Bonn, Germany \and Harvard-Smithsonian Center for Astrophysics, 60 Garden Street MS 72, Cambridge, MA 02138, U.S.A.,  \email{jforbrich@cfa.harvard.edu} \and Instituto de Radioastronom\'ia Milim\'etrica, Avenida Divina Pastora 7, Local 20, 18012 Granada, Spain \and Institut de Radioastronomie Millim\'etrique, Rue de la Piscine, Saint Martin d'H\`eres, France}

   \date{Received; accepted}

    \abstract{The Goldreich-Kylafis (GK) effect causes certain molecular line emission to be weakly linearly polarized, e.g., in the presence of a magnetic field. Compared to polarized dust emission, the GK effect has the potential to yield additional information along the line of sight through its dependence on velocity in the line profile.}{Our goal was to detect polarized molecular line emission toward the DR21(OH), W3OH/H$_2$O, G34.3+0.2, and UYSO~1 dense molecular cloud cores in transitions of rare CO isotopologues and CS. The feasibility of such observations had to be established by studying the influence of polarized sidelobes, e.g., in the presence of extended emission in the surroundings of compact sources.}{The observations were carried out with the IRAM 30m telescope employing the correlation polarimeter XPOL and using two orthogonally polarized receivers. We produced beam maps to investigate instrumental polarization.}{While in nearly all transitions toward all sources a polarized signal is found, its degree of polarization only in one case surpasses the polarization that can be expected due to instrumental effects. It is shown that any emission in the polarized sidelobes of the system can produce instrumental polarization, even if the source is unpolarized. Tentative evidence for astronomically polarized line emission with $p_{\rm L}\lesssim1.5\%$ was found in the CS(2-1) line toward G34.3+0.2.}{} 


   \maketitle

\section{Introduction}
\label{intro}

Currently, two competing theoretical scenarios are being discussed that aim to explain the formation of low-mass (i.e. solar-like) stars. The 'standard' model (e.g. \citealp{mou99}) invokes strong magnetic fields and ambipolar diffusion for determining the star formation time scale. This paradigm is challenged by the picture of turbulence-controlled star formation (e.g. \citealp{bal06}) in which magnetic fields play no significant role. In the latter picture, molecular clouds are dynamically controlled by turbulence and short-lived; star formation starts from transient density enhancements. It remains unclear, however, how turbulence on different spatial scales is constantly replenished. On the other hand, whether or not the star formation time scale in the ambipolar-diffusion picture is compatible with observations is a matter of debate (e.g. \citealp{mou06}). A major difference between these two paradigms concerns the earliest stages of star formation before a central protostar has formed. It is also possible that both scenarios may apply for different circumstances. Measurements of magnetic fields in star-forming regions will help refining and deciding between the two paradigms. For instance, results from interferometric dust polarimetry toward the protostellar system NGC~1333\,IRAS\,4A show an hourglass-shaped magnetic field morphology \citep{gir06}, suggesting that magnetic fields are not negligible, at least not in this region.

There are three different possibilities for detecting magnetic fields in molecular clouds, first, observations of the frequency shifts the Zeeman effect causes between right and left circularly polarized components of a spectral line signal, second, measurements of the polarized thermal dust continuum emission, and third, linear polarization of molecular rotational emission lines \citep[e.g.][]{cru03}. These methods are sensitive to different projections of the magnetic field vector (parallel to the line of sight for Zeeman observations and perpendicular to it for polarization measurements). \citet{hou05} describe a method that uses a combination of submillimeter polarimetry and measurements of ion-to-neutral molecular line width ratios to reconstruct the three-dimensional magnetic field distribution.

Weak linear polarization in rotational emission lines of molecules was first predicted by \citet{gok81,gok82}. Additionally, \citet{kys83} and \citet{kyl83b} also predict polarized absorption lines. Already a very weak magnetic field causes the rotational levels to split into magnetic sublevels. Unequal populations of the different sublevels lead to partial linear polarization. This is the case when there is a gradient in the line optical depth, e.g. due to a velocity gradient or an anisotropic radiation field. In this model, the polarization percentage depends on several factors: the degree of anisotropy, the ratio C/A of the collisional excitation rate to the radiative decay rate, i.e. the ratio of the density compared to the critical density of the transition, the optical depth of the line, and the angle between the line of sight, the magnetic field, and an assumed axis of symmetry of the velocity field. The maximum fractional polarization is on the order of a few percent and is expected for C/A$\approx 1$ -- i.e. a density close to the critical density of the transition -- and $\tau \approx 1$. Since the optical depth varies across a line profile, an optical depth of order unity is usually found only in certain parts of a spectrum.

Detailed predictions of the Goldreich-Kylafis (GK) effect and its observables are model-dependent. Trying to predict the polarization percentage, \citet{gok81,gok82} used a large velocity gradient (LVG) model with two molecular rotational levels, resulting in values of up to 14\%. An extension of this model to include more levels, developed by \citet{dew84}, resulted in lower polarizations of about 7\%.
\citet{mor85} discussed line polarization in expanding circumstellar envelopes -- in fact the environment in which the effect was first observed later on. \citet{lis88}, abandoning the LVG approximation, concluded that polarizations of a few percent should be observable. According to them, polarization should be greatest for molecules that have large permanent electric dipole moments, $\mu$, otherwise collisions govern the level populations and, if isotropic, tend to equalize them. Therefore, polarized CS emission would be more likely observable than from CO ($\mu=2.0$~D vs. 0.1~D), even though CS emission may be fainter. \citet{kyl83} and, e.g., \citet{lis88} note that linear polarization is expected to be stronger at the {\sl edge} of molecular clouds rather than at their centers due to a more anisotropic distribution of optical depth and because the high gas densities at the source centers tend to equalize the sublevel populations by isotropic collisions. We will see in Section~\ref{simusect} that such observations are technically very difficult with a single-dish telescope because it is very likely that strong emission from a cloud's core may fall onto polarized instrumental sidelobes.

The observed polarization direction is linked to the orientation of the magnetic field projected onto the plane of the sky. According to \citet{kyl83}, the polarization is  either perpendicular or parallel to the projection of the magnetic field lines in case of simple velocity fields (1D, 2D, or axisymmetric), depending on the properties of the velocity field. This holds even for quite low field strengths ($B \approx 1 \mu$G). 

While polarization observations of thermal dust emission have been possible for many years, the low polarization levels expected for molecular lines were discovered only recently. Following earlier attempts toward young stellar objects \citep{wan83,bar87,gle97a}, the first observation of linearly polarized molecular line emission was made by \citet{gle97b} toward the envelope of the evolved star IRC +10216 in CS(2-1). They measured polarizations of $p=5.1 \pm 1.5$\%, as of now the highest polarization interpreted as due to the Goldreich-Kylafis effect. Then, \citet{gre99} observed two molecular clouds belonging to the region around the Galactic center and the ``2 pc ring'' surrounding it in the CO(2-1) and CO(3-2) lines and found polarizations at a level of 0.5\%$-$2\% . The Goldreich-Kylafis effect has also been observed in interferometric observations toward the outflow of the young stellar system NGC~1333~IRAS~4A \citep{gir99}, as well as in the outflow of NGC~2024~FIR~5 in single-dish observations \citep{gre01}. Also, gas flows near the galactic center were studied \citep{ghd02}. \citet{lai03} measured polarized continuum and CO line emission in the star-forming region DR21(OH), confirming the prediction that the two polarizations are either parallel or orthogonal to each other. \citet{ccw05} find that toward the same source, the linear polarization of the CO(1-0) transition is perpendicular to that of the CO(2-1) transition, an unexpected result which they explain by anisotropic excitation. In subsequent BIMA studies, weak linear polarization of the CO(2-1) line was found in G30.79~FIR~10 and NGC2071IR \citep{cor06,cor06b}, as well as in CO(1-0) toward the massive-star--forming region G34.4+0.23~MM \citep{cor08}.

When compared to dust polarization, line observations have the advantage of containing velocity information. Continuum and line polarization observations thus are complementary, covering a wide parameter space of density and temperature. A complication of polarization measurements of extended sources is the presence of polarized instrumental sidelobes. They can easily generate false weak polarization signals, particularly in line observations where strong transitions of abundant molecules are measured. For our observations, we therefore also studied the influence of polarized sidelobes on our measurements. In Section~\ref{obs}, we describe the observations carried out, before presenting and discussing the results in Sections~\ref{res} and \ref{dis}. We conclude with a summary in Section~\ref{sum}.

\section{Observations and data analysis}
\label{obs}

\begin{figure*}
\centering
\includegraphics*[width=0.9\linewidth]{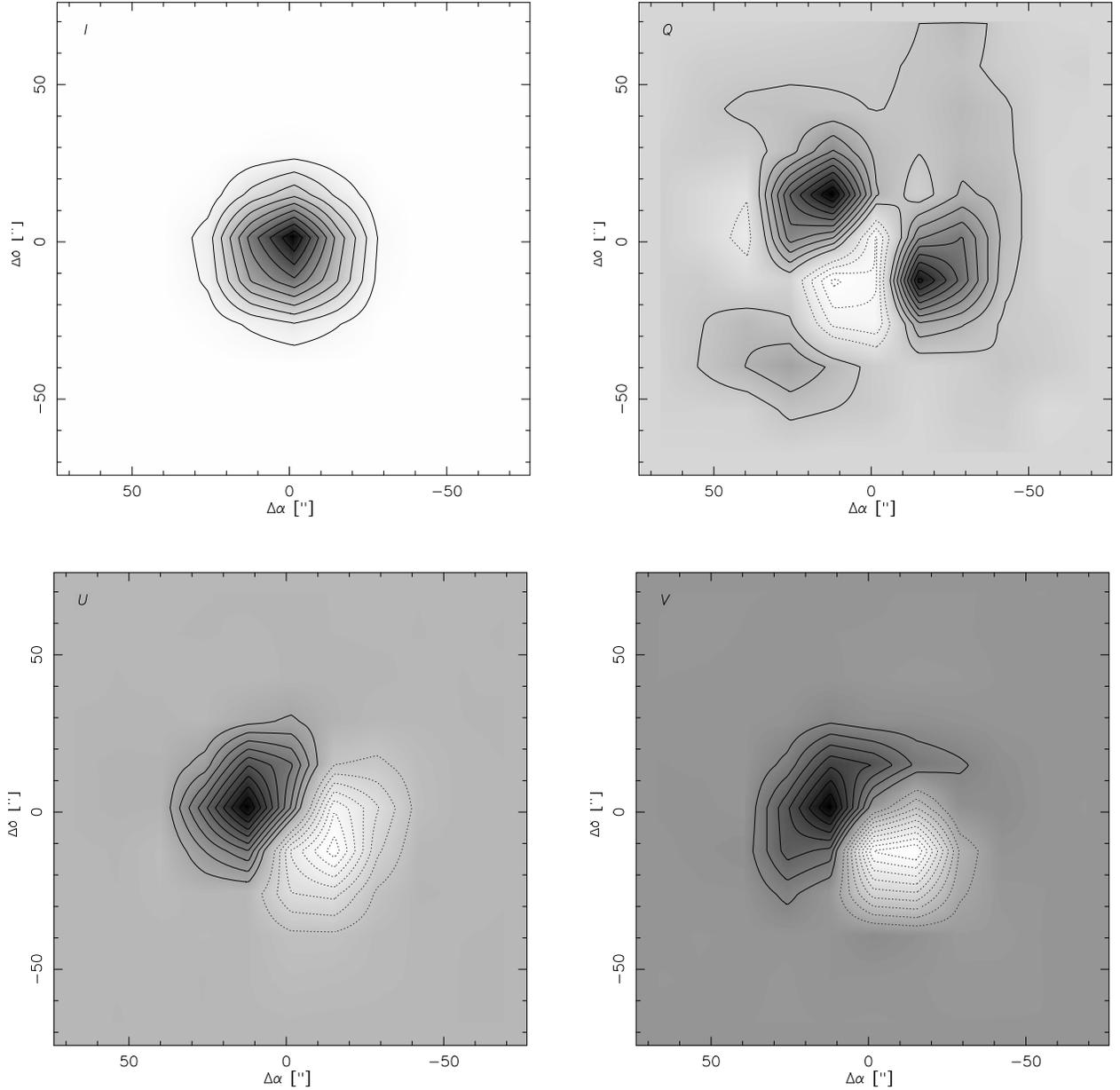}
\caption[Beam maps of the 30m telescope (in Stokes-$I$, $Q$, $U$, and $V$)]{Beam maps of the 30m telescope (Stokes-$I$, $Q$, $U$, and $V$), acquired by observations of Venus and used in the simulations. The contour lines delineate multiples of 10\% peak intensity (without zero), negative values are marked as dotted lines. The maxima of the Stokes-$Q$,-$U$, and -$V$ maps, in relation to the peak in Stokes-$I$, are at 2.1\%, 1.3\%, and 0.8\%, respectively (not generally at the image center).}
\label{beamfig}
\end{figure*}

\begin{table}
\caption{Source coordinates}
\label{srctab}     
\centering                
\begin{tabular}{lrrrr}  
\hline\hline              
Source    & RA/Dec (J2000.0) & v$_{\rm lsr}$  & d\\ 
          &                  & [km\,s$^{-1}$] & [kpc]\\
\hline 
DR21(OH)  & 20h39m00.9s   +42$^\circ22'48\farcs0$   &  +3  & 3$^{\rm a}$\\
G34.3+0.2 & 18h53m18.6s   +01$^\circ15'00\farcs0$   & +58  & 3.8$^{\rm b}$\\
W3OH/H$_2$O&02h27m03.9s   +61$^\circ52'25\farcs0$   &--48  & 2.0$^{\rm c}$\\
UYSO\,1   & 07h05m10.85s --12$^\circ19'02\farcs4$   & +12  & 1.0$^{\rm d}$\\
\hline                                   

\multicolumn{4}{l}{References for the distances:} \\
\multicolumn{4}{l}{a) \citet{cam82}, b) \citet{rei85}} \\
\multicolumn{4}{l}{c) \citet{hac06}, d) \citet{for04}} \\
\end{tabular}
\end{table}

\begin{table}
\caption{Molecular transitions, dates, XPOL integration times, and hour angle ranges (not covered continuously)} 
\label{obstab}     
\centering                
\begin{tabular}{lllrrr}  
\hline\hline              
Source & Transition & Date & t$_{\rm int}$ & HA$_{\rm start}$ & HA$_{\rm end}$\\ 
       &            &      & [min]         & [h]              & [h]   \\
\hline 
DR21(OH)  & $^{13}$CO(1-0) & Apr 23 & 144  & --1.10 &   6.61 \\ 
DR21(OH)  & C$^{18}$O(1-0) & Apr 22 &  62  & --5.32 & --2.05 \\ 
DR21(OH)  & CS(2-1)        & Apr 22/23 &  72  & --2.26 &   8.00 \\ 
G34.3+0.2 & $^{13}$CO(1-0) & Apr 23 &  80  &	1.86 &   4.29 \\ 
G34.3+0.2 & C$^{18}$O(1-0) & Apr 22 &  64  &	0.27 &   2.04 \\ 
G34.3+0.2 & CS(2-1)        & Apr 23 &  64  & --2.73 & --1.06 \\ 
W3OH/H$_2$O    & $^{13}$CO(1-0) & Apr 22 &  64  &	3.24 &   5.00 \\ 
W3OH/H$_2$O    & C$^{18}$O(1-0) & Apr 22 &  64  & --4.96 & --2.82 \\ 
W3OH/H$_2$O    & CS(2-1)        & Apr 22 &  66  & --1.54 &   0.40 \\ 
W3OH/H$_2$O    & C$^{34}$S(2-1) & Apr 23 &  88  &	4.46 &   6.97 \\ 
UYSO\,1   & $^{13}$CO(1-0) & Apr 23 &  72  & --3.14 & --0.68 \\ 

\hline                                   

\end{tabular}
\end{table}

We used the XPOL polarimeter at the IRAM 30m telescope \citep{thu08} to search for linearly polarized molecular line emission. XPOL operates at the intermediate frequency of the telescope, and uses its two 3mm receivers (A/B100) whose orthogonally linearly polarized signals are auto-correlated  (to give Stokes $I$ and $Q$ in the Nasmyth reference frame) and cross-correlated (to give Stokes $U$ and $V$ in the Nasmyth reference frame) in the digital backend VESPA. We observed with a band width of 105~MHz and a channel width of 39~kHz, leading to a velocity resolution of about 0.1\,km\,s$^{-1}$. The phase difference between the receivers is calibrated before every observation for each spectral channel by looking at a strong linearly polarized continuum source in the receiver cabin.  While the on--axis instrumental linear and circular polarization was low (approx. 0.7\%), significantly stronger sidelobes exist in Stokes $Q$ and $U$, mainly due to residual receiver misalignment.  Beam maps of these Stokes parameters, obtained on August 30, 2004 (see Fig.~\ref{beamfig} and also \citealp{thu08} for beam maps after improved beam alignment and further minimized cross-polarization), have been used to estimate the instrumental linear polarization due to these sidelobes (Section~\ref{simusect}). The observations were carried out on April 22 and 23, 2005, under good weather conditions, with system temperatures between 96~K and 217~K. For removal of the atmospheric contribution to the total power signal, the wobbler was used with a throw of $200''$ in azimuth. 

We observed the star-forming region DR21(OH) and a hot molecular core near the ultra-compact \HII region W3OH/H$_2$O, as well as the cometary \HII region G34.3+0.2 mainly in the $^{13}$CO(1-0), C$^{18}$O(1-0), and CS(2-1) lines in order to detect linear polarization due to the Goldreich-Kylafis effect. In addition, the candidate massive protostar UYSO~1 \citep{for04} was observed in the $^{13}$CO(1-0) line. The sources were mainly selected for their bright line emission and their compactness compared to the telescope beam of, e.g., $22''$ FWHM for the CO(1-0) line.

An earlier observing run was conducted in April 2004.
In-between the two observing campaigns, the alignment of the two receivers had been considerably improved, reducing instrumental polarization. Consequently, we here discuss only the April 2005 data. Details about the observations are listed in Tables~\ref{srctab} and~\ref{obstab}. The pointing was checked approximately hourly while the focus was checked about every two hours. The standard deviation of the pointing corrections throughout this observing run is $2\farcs7$.

While previous polarimetric observations of DR21(OH) in $^{12}$CO \citep{lai03} are sensitive only to the outermost material or foreground material along the line of sight (although extended emission is filtered out by the use of an interferometer), our choice of CO isotopologues with a lower optical depth should improve the comparability with dust polarization measurements. With CS, the same region is also traced using a molecule with a larger dipole moment (2.0 vs. 0.1~Debye) and thus higher critical density (by about two orders of magnitude).  The maximum linear polarization in the CO(2-1) line reported by \citet{lai03} is $2.4 \pm 0.6$~\%.

We analyzed the data with the XPOL software version 3.0, developed in the CLASS environment of the GILDAS software\footnote{http://www.iram.fr/IRAMFR/GILDAS}. In the data reduction, an elevation-dependent part of the instrumental polarization is taken into account, as determined from observations of Venus at different elevations. This leads to a correction for horizontal polarization in Nasmyth coordinates of typically $<1$~\% (for elevations between $13^\circ$ and $62^\circ$).

In order to discuss the role of extended emission in our data, on-the-fly maps of source neighbourhoods were obtained on August 16--18, 2006 at the 30m telescope. Mapping was performed in molecular line transitions which were also used in the previous polarimetric observations. Two molecular lines were observed simultaneously toward each of the three sources W3OH/H$_2$O, G34.3+0.2, and DR21(OH). In case of W3OH/H$_2$O, the mapping was done in C$^{18}$O(1-0) and $^{13}$CO(1-0) while for the latter two sources the transitions $^{13}$CO(1-0) and CS(2-1) were chosen.

\section{Results}
\label{res}

\begin{figure}
  \includegraphics*[width=8.9cm, bb=20 35 395 720]{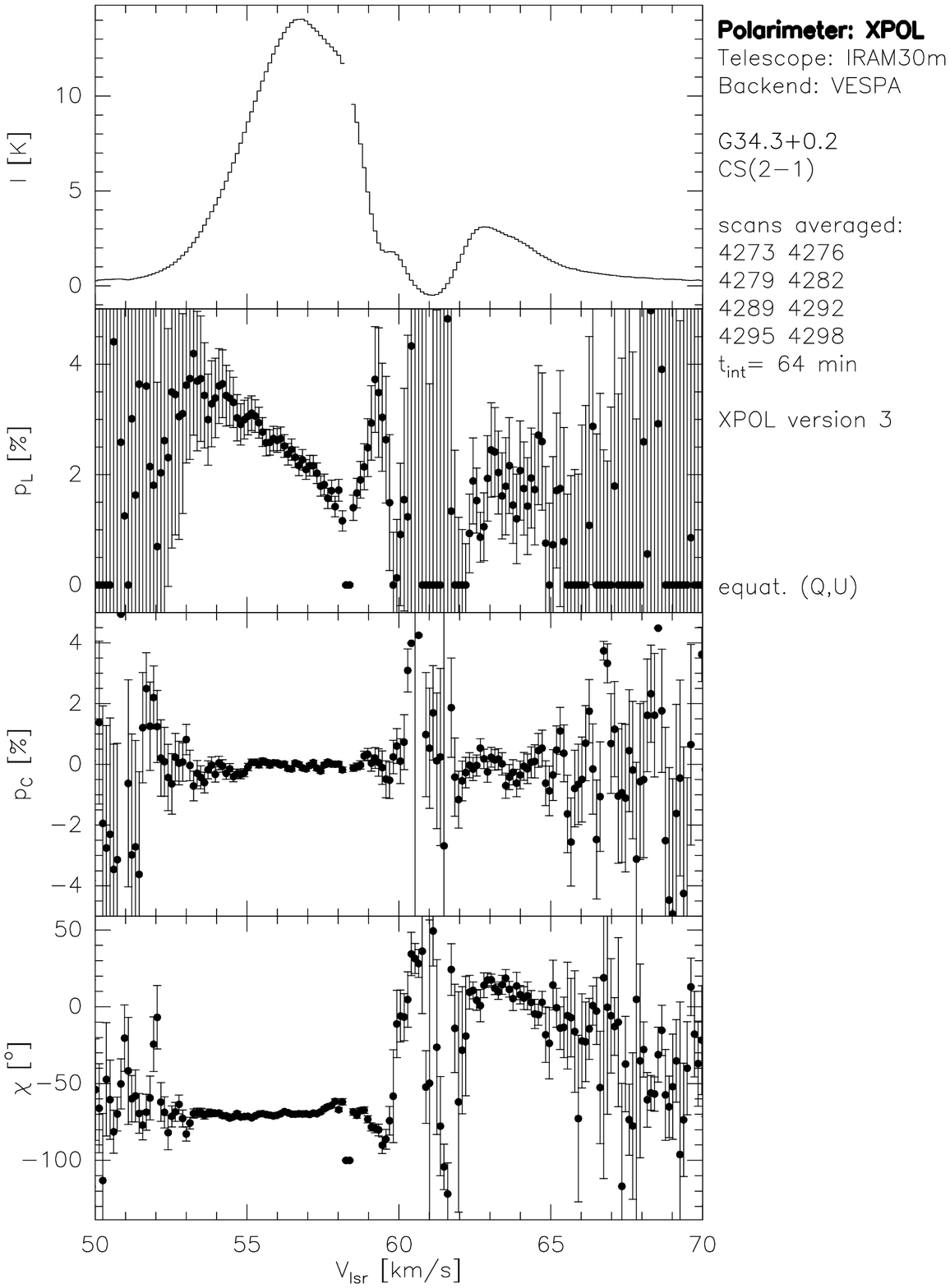}
  \caption{G34.3+0.2 in CS(2-1): Resulting linear ($p_{\rm L}$, $\chi$) as well as circular ($p_{\rm C}$) polarizations [\%, $^\circ$]. The spectrum in the uppermost panel is in $T_A^*$. A baseline was subtracted and a few single-channel spikes were blanked. See also Fig.~\ref{figresults_onl}. The error bars represent $1\sigma$ baseline noise from the four Stokes spectra propagated throughout the analysis.}
  \label{g34.3_CS}
\end{figure}

\begin{figure*}
\centering
\includegraphics*[width=14cm]{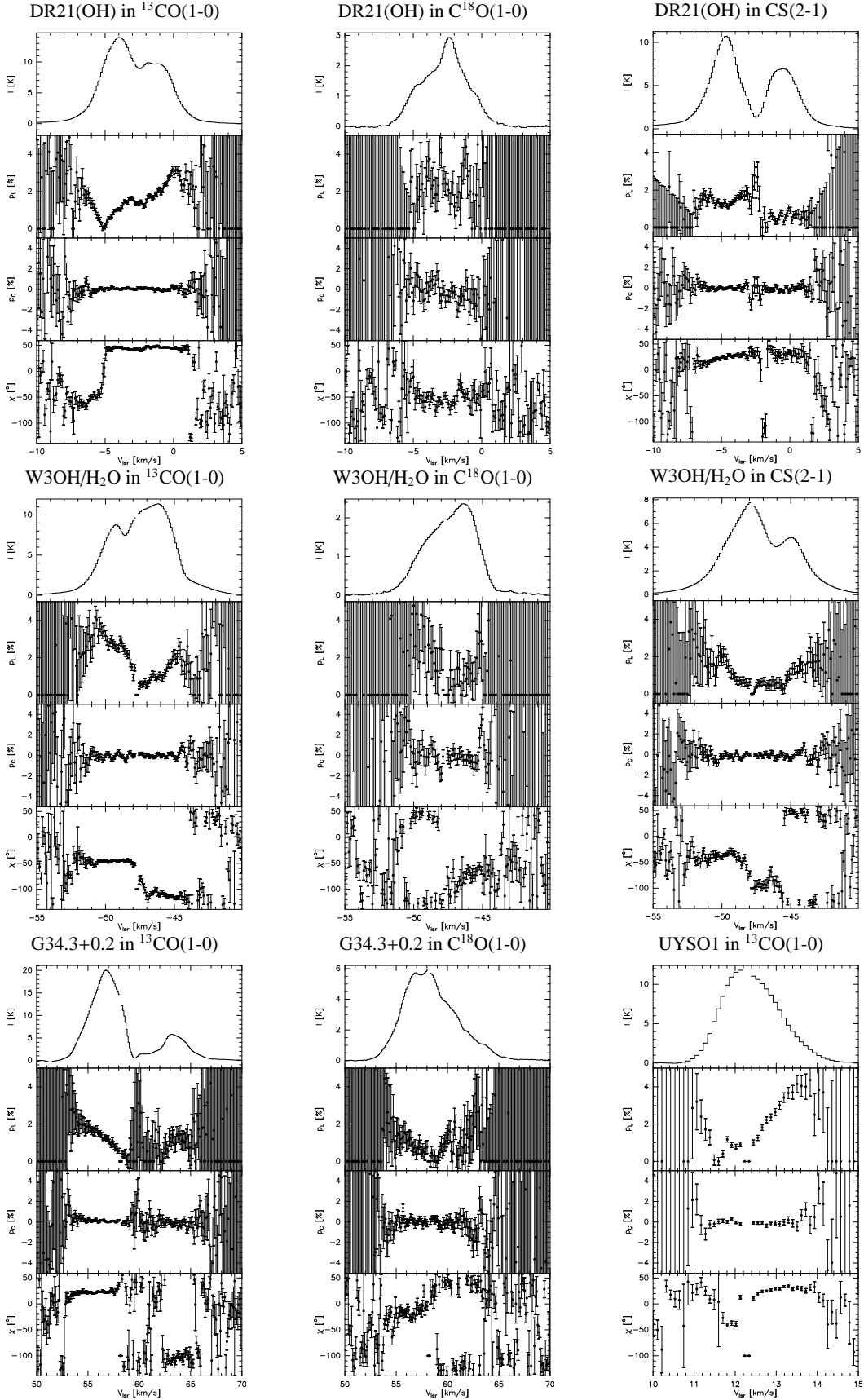}
\caption[Resulting linear ($p_{\rm L}$, $\chi$) as well as circular ($p_{\rm C}$) polarizations]{Resulting linear ($p_{\rm L}$, $\chi$) as well as circular ($p_{\rm C}$) polarizations [\%, $^\circ$] (see also Fig.~\ref{g34.3_CS}).  The spectra in the uppermost panels are in $T_A^*$. A baseline was subtracted and a few single-channel spikes were blanked. Sources and transitions are indicated. The error bars represent $1\sigma$ baseline noise from the four Stokes spectra propagated throughout the analysis.}
\label{figresults_onl}
\end{figure*}

Based on the Stokes spectra, the fractional linear polarization $p_{\rm L} = \sqrt{(Q^2+U^2)}/I$, the circular polarization $p_{\rm C} = V/I$ and the position angle of the linear polarization $\chi = {\rm arctan}(U/Q)/2$ (E from N) are calculated. 
The derived linear and circular polarizations as well as the polarization angles, are shown in Fig.~\ref{g34.3_CS}, where we single out the most interesting measurement, G34.3+0.2 in the CS(2-1) line, and in Fig.~\ref{figresults_onl}, respectively. In all cases, weak linear polarization, varying across the line profiles, is detected while the circular polarization is zero within the uncertainty of the measurement across the entire line profiles. It is interesting to note that the linear polarization varies significantly with line velocity in all sources. The fractional linear polarizations are below 4\%, and this value is reached only in few cases, e.g. for G34.3+0.2 in CS(2-1). 
Toward DR21(OH), linear polarizations of up to 3\% were found in the $^{13}$CO and CS lines, but the dependency on velocity is quite different, with notably the $^{13}$CO polarization highest in the redshifted line wing. The C$^{18}$O data have a low signal-to-noise ratio. Also in W3OH/H$_2$O, only the $^{13}$CO and the CS data are of good quality. As toward DR21(OH), the polarization in $^{13}$CO is slightly higher than in CS, at about 4\%. The dependency on velocity is more similar here, but still significantly different. In the case of G34.2+0.2, polarization is detected in all three lines, most notably in the CS, but also in the $^{13}$CO and even the C$^{18}$O lines. The polarization in CS is slightly stronger than in $^{13}$CO, but the dependency on velocity is remarkably similar (as is the line profile). The observed polarization appears to be systematically higher in the line wings, even when taking into account the lower signal to noise ratio. In at least one case -- G34.3+0.2 in CS(2-1) -- the polarization also rises toward the line center, although at a low significance.
The polarization angles mostly have well-defined values, smoothly varying or even nearly constant across parts of the line profile. This does not necessarily have to be a property of the sources, though (see discussion of instrumental effects below). 

\subsection{Simulating the instrumental response}
\label{simusect}

There is always an instrumental contribution to the observed linear polarization. A clear sign of instrumental polarization is the leakage of power from Stokes-$I$ to other Stokes parameters, especially to Stokes-$Q$ and $U$.

While in some of our measurements the Stokes-$Q$ spectra resemble scaled copies of the Stokes-$I$ spectra at first sight, possibly a sign of this leakage effect, the correspondence is not exact, leading to considerable residuals when trying to correct for this by subtraction. This non-correspondence can be explained by velocity structure of the source which would lead to the observation of slightly different source parts in the different Stokes parameters. One consequence of this polarization leakage can be a mostly constant polarization angle $\chi$: If the Stokes-$Q$ and -$U$ spectra are scaled copies of the Stokes-$I$ spectrum to some degree, their ratio and, thus, $\chi$ will tend to stay constant over the spectrum. 

In order to determine the order of magnitude of the instrumental contribution to the polarized signal more quantitatively, including the role of extended emission, we have attempted to simulate the linear polarization based on real source geometries but assuming that the sources are \textsl{not} intrinsically polarized. How the telescope beam looks like in the different Stokes parameters is best probed by observations of bright unpolarized sources. We acquired such beam maps in all four Stokes parameters by continuum observations of Venus on August 30, 2004, at a wavelength of 3~mm. 
At this time, Venus had an apparent size of $20\farcs9$. The alignment of the receivers has not been changed between this date and our polarization observations in April 2005, although the alignment may have drifted with time. The signal-to-noise ratio in the Stokes-$I$ map is $\sim500$.  The linear polarization at the (0,0) position (i.e. the map center) is 0.74~\% and stays $<1$~\% within reasonable pointing accuracies. These beam maps contain information on the instrumental polarization, for instance the effects due to `beam squint', i.e., due to the fact that the two receivers are pointing to slightly different positions on the sky leading to residual polarization in the sky area covered by the sidelobes. Therefore, the presence of extended emission can significantly alter the measurements, as can the presence of point-like emission if the telescope pointing is imperfect. Convolving the measured polarized beam profiles with the observed source geometries, assuming no intrinsic polarization thus yields a good estimate of the instrumental limits when observing low percentages of linear polarization, although a detailed error analysis of the simulation would require several independently measured beam maps, which was not attempted in our study. 

\begin{figure*}
\centering
\includegraphics*[width=0.85\linewidth]{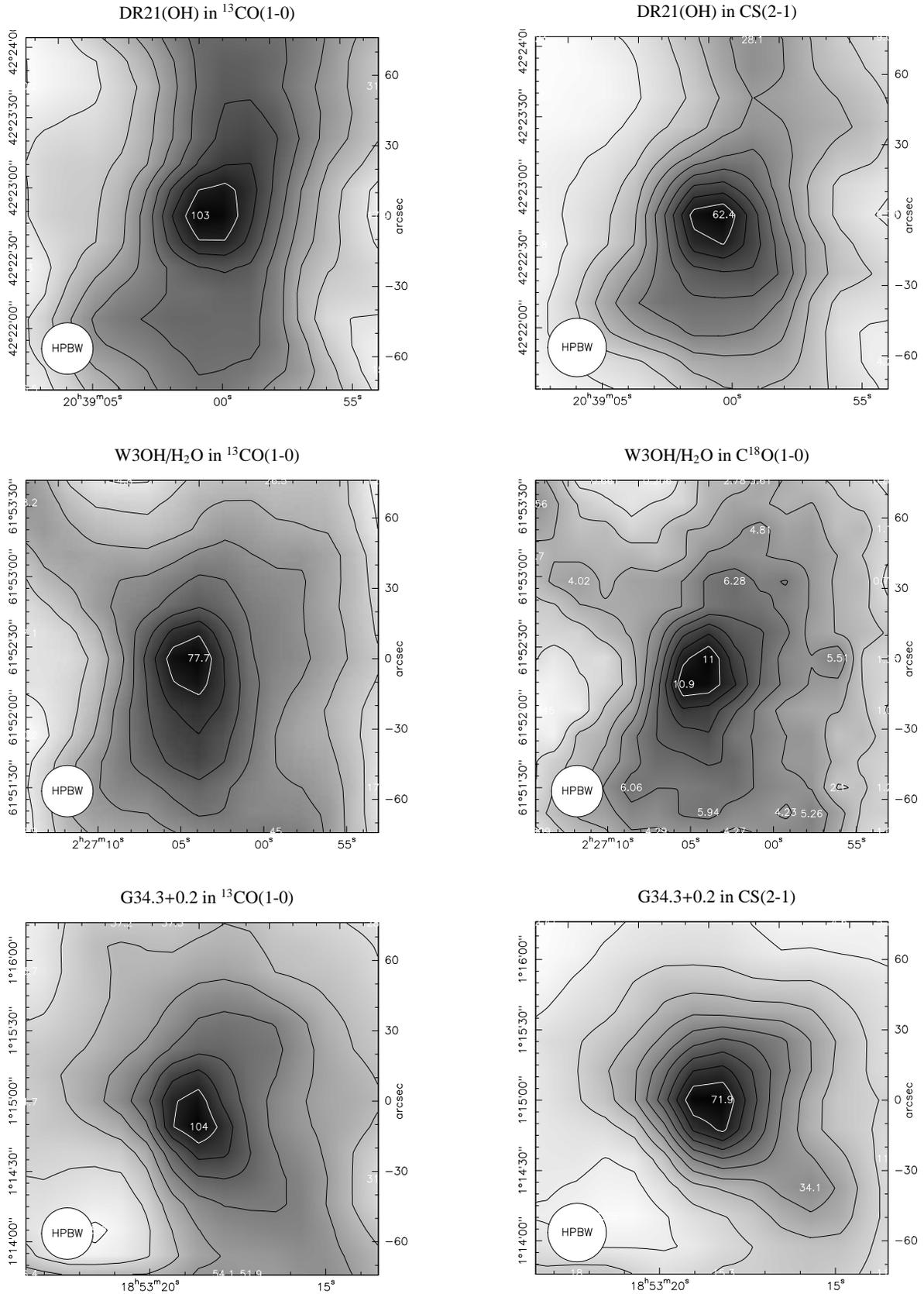}
\caption[On-the-fly maps of DR21(OH), W3OH/H$_2$O, and G34.3+0.2]{On-the-fly maps of  DR21(OH), W3OH/H$_2$O, and G34.3+0.2, obtained with the IRAM 30m telescope, in two molecular transitions per source (as labeled). The contour lines delineate multiples of 10\% of the map maximum which is indicated (the 90\% contour line is shown in white for clarity). The XPOL observations were pointed at the map centers. Map unit is K~km\,s$^{-1}$ ($T_A^*$ integrated over the entire line profile), coordinates are in J2000.0. }
\label{otffig}
\end{figure*}

As noted above, the maps of the source geometries on the order of several beam sizes were obtained by on-the-fly mapping in two molecular transitions for DR21(OH), G34.3+0.2, and W3OH/H$_2$O (Fig.~\ref{otffig}). While the emission is well concentrated in the main source peaks there is (of course) still considerable emission in the surroundings. 

In order to simulate the dependency of the linear polarization $p_{\rm L}$ on frequency (i.e. velocity), first a series of channel maps of 0.5\,km\,s$^{-1}$ width was computed. After resampling to the same velocity resolution, these were then convolved with the polarization beam maps in Stokes $Q$, $U$, and $V$ of the 30m telescope in order to simulate a polarization observation. Since the observed maps are already convolved with the telescope beam, this is not entirely accurate, but the additional smoothing due to a second convolution with the beam should not interfere with the estimates of instrumental effects on linear polarization that we seek to derive. Also, a simple convolution of source and model beam does not take into account the rotation of the source structure with regard to the Nasmyth reference system which necessarily occurs in an observation covering an extended hour-angle range (the bulk of instrumental polarization comes from the receiver cabin, hence the Nasmyth focus). Subsequently, maps of simulated linear polarization were computed, and the value of $p_{\rm L}$ at the map center was taken as the simulated linear polarization. In general, the simulated polarization increases with distance from the map center because the main source emission then falls onto the polarized sidelobes. We note that within $\pm10''$ of the map center, more than enough to accomodate our pointing uncertainty (see above), the simulated instrumental polarization varies by only about $\pm0.2\%$. However, we note that if the telescope is pointed further away from the unpolarized emission peak, the simulated linear polarization rises to significantly higher levels, reaching values of 10\% and more, presumably when the main peak is picked up by polarized sidelobes.

\begin{figure*}
\centering
\includegraphics*[width=0.85\linewidth]{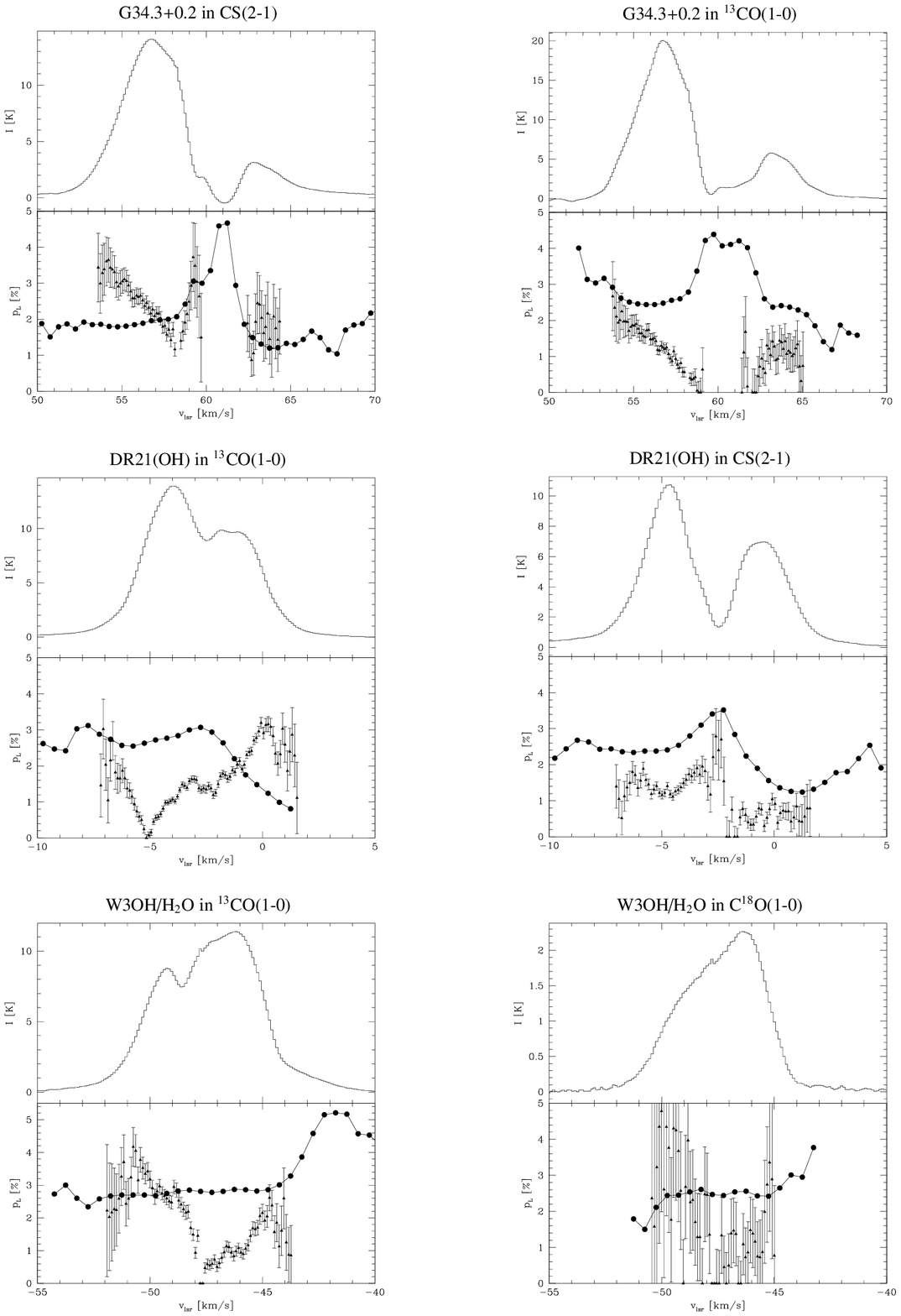}
\caption[Simulated linear polarization after convolving the measured maps of the line integrated source environments with the Stokes beam maps]{Upper panels: line profiles, lower panels: simulated linear polarization after convolving the measured integrated intensity maps of the source environments with the Stokes beam maps (big points connected by a line, with actual data shown as triangles). For clarity, only measured datapoints with errors $<\pm1\%$ are shown. For the full observational data, see Figs.~\ref{g34.3_CS} and \ref{figresults_onl}.}
\label{simufig}
\end{figure*}

The resulting simulated spectra for G34.3+0.2 in the CS(2-1) line, as well as for the other sources, are shown in Fig.~\ref{simufig}.
In all cases studied here, it is important to note that the simulated linear polarizations assuming intrinsically unpolarized sources have values that are of the same order of magnitude as the actually observed ones. At typically $>2$\%, the simulated linear polarization is considerably higher than the value of 0.74\% from the beam map alone (see above and Fig.~\ref{simufig}), clearly demonstrating the influence of extended emission on the result of the measurement. For most sources and transitions, there is no apparent correlation of simulated and measured polarization. We note again that the simulations do not have the aim to exactly reproduce the measured polarizations but to estimate the order of magnitude of the contamination of the Stokes spectra by polarized sidelobes.

\section{Discussion}
\label{dis}

\subsection{Comparison to previous results}

Since DR21(OH) was successfully searched before for the GK effect \citep{lai03}, though with an interferometer and in $^{12}$CO(1-0), it is interesting to compare the results. With the 30m telescope, we can detect polarized line emission on much larger spatial scales. Looking at previous results of millimeter continuum and heterodyne polarimetry, the assumption that magnetic fields are detectable on these large spatial scales seems reasonable \citep[e.g.][]{war00}, even though it is important to keep in mind that such observations are not resolving smaller angular scales. The opposite problem occurs in observations using an interferometer without zero-spacing correction yielding information on large-scale emission. \citet{lai03} find a maximum degree of linear polarization of $2.4 \pm 0.6$~\% in the $^{12}$CO(1-0) line and a polarization angle rapidly varying with velocity. How these results based on the much more optically thick $^{12}$CO(1-0) line compare in detail to our single-dish observations of less optically thick isotopologues, when filtering out large-scale emission by the use of an interferometer, remains unclear. We note, however, that the estimated instrumental polarization in our measurements is larger than the polarization reported by \citet{lai03}, possibly preventing us from detecting astronomical polarization. In our data, a comparison is further hampered by significant variations between different scans of the Stokes-$I$ line profiles, mainly in $^{13}$CO. Since the interferometric $^{12}$CO data reported by \citet{lai03} clearly show two emission peaks separated by $8''$, i.e. about a third of the beam size in our observations, this may be due to marginally resolved source structure. In this case, the varying line profiles could be explained by small pointing inaccuracies (the standard deviation of our pointing corrections was $2\farcs7$, see above). The same effect, though to a lesser degree, is also observed toward W3OH/H$_2$O, again mainly in $^{13}$CO.

\subsection{G34.3+0.2: intrinsically polarized?}

The measurements yielded values above the simulated polarizations only in a few cases. The averaged simulated linear polarization is lowest for G34.3+0.2 in the CS(2-1) line, arguably the most centrally concentrated source in this sample (Fig.~\ref{otffig}). In the blueshifted line wing, measured polarizations of up to 3.6\% contrast with simulated polarizations of slightly below 2\%. If we take the simulation as an estimate for the instrumental polarization, then after its subtraction the intrinsic polarization reaches values of up to $\sim$1.6\%. The measurements are also above the simulated values for DR21(OH) and W3OH/H$_2$O in the $^{13}$CO(1-0) transition, however these two datasets are less reliable due to the above-mentioned time-variable line profiles.

Thus, G34.3+0.2 emerges as a tentative detection of intrinsically polarized radiation within our sample. Additionally, the source appears to be sufficiently unresolved so that minor pointing inaccuracies do not affect the line profiles in the individual scans in both the CS(2-1) and $^{13}$CO(1-0) lines, allowing a comparison of the different transitions.

The spectra of G34.3+0.2 show two peaks in the $^{13}$CO and CS lines, apparently due to self-absorption at the line center (see the discussion of optical depth at the end of this subsection). In the C$^{18}$O line, only a single, broad line profile is seen. The observed linear polarization is strongest for the CS and $^{13}$CO lines with values of up to $\sim4\%$ while in C$^{18}$O, values $\lesssim2\%$ were measured. The velocity dependence of the linear polarization is remarkably similar especially for $^{13}$CO(1-0) and CS(2-1). The most notable feature is that the polarization is highest in the blue-shifted line wing of the main peak, steadily falling off toward the red-shifted line wing. Interestingly, the offset between the polarization angles $\chi$ of the $^{13}$CO and CS lines (as a function of velocity) is roughly 90$^\circ$ across the main peak. The situation is less clear for the weaker peak. This differs from W3OH/H$_2$O where there is no significant offset between the polarization angles in the $^{13}$CO(1-0) and CS(2-1) lines in nearly the entire spectrum. However, this source probably does not show astronomical polarization (Fig.~\ref{simufig}). Differences of $\pm90^\circ$ are reminiscent of the results of \citet{ccw05} who, as noted above, observed such a difference between the CO(1-0) and CO(2-1) transitions.

\begin{figure}
\centering
\includegraphics*[width=9.3cm, bb=60 130 560 670]{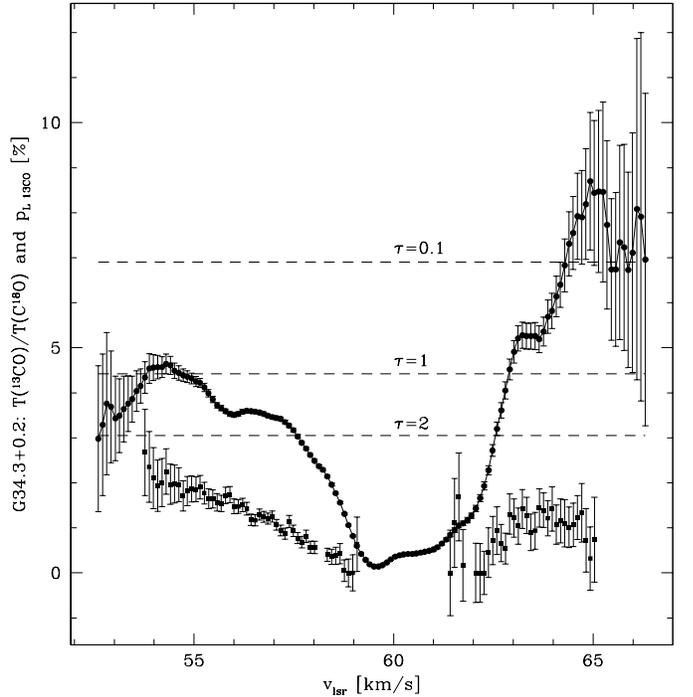}
\caption[Line ratio $^{13}$CO(1-0)/C$^{18}$O(1-0) for G34.3+0.2 (where both lines have $T_a>0.1$~K) with inferred optical depth $\tau$ of the $^{13}$CO(1-0) transition indicated]{Line ratio $^{13}$CO(1-0)/C$^{18}$O(1-0) for G34.3+0.2 (where both lines have $T_a>0.1$~K; shown as connected points) with inferred optical depth $\tau$ of the $^{13}$CO(1-0) transition indicated (see text). Squares with error bars denote the observed linear polarization in $^{13}$CO (from Fig.~\ref{simufig}).}
\label{g34.3_tau}
\end{figure}

Since the Goldreich-Kylafis effect is expected to be most prominent for $\tau\sim1$, it is of interest to estimate the optical depths of the transitions in question.
The measurement of both the $^{13}$CO(1-0) and C$^{18}$O(1-0) isotopologue lines during this project allows us to estimate the optical depth of the $^{13}$CO(1-0) transition from the $^{13}$CO/C$^{18}$O line ratios based on a number of assumptions. Among others, the beam filling factors are assumed to be the same, and the ratio of optical depths is assumed to be $\tau_{13}=7.3\tau_{18}$ \citep[e.g.][]{wir94}.  Fig.~\ref{g34.3_tau} shows the observed $^{13}$CO/C$^{18}$O line ratios for G34.3+0.2, together with indications of the inferred optical depth of the $^{13}$CO transition as well as its observed linear polarization. The optical depth is indeed close to unity throughout most of the blue-shifted line wing where the observed polarization is highest. Optical depth and polarization appear to be anticorrelated in that range. In the redshifted line wing, where the optical depth falls off steeply, similar degrees of polarization are observed, although the detections are less significant. 

\section{Summary}
\label{sum}

We present results of a search for weak linear polarization in molecular line emission toward protostellar sources, as expected if the Goldreich-Kylafis effect occurs. The project was carried out with the correlation polarimeter XPOL at the IRAM 30m telescope in different molecular line transitions. In order to interpret the spectra, we simulated the instrumental response by convolving maps of the sources integrated in small velocity intervals with the telescope beam in all Stokes parameters. In this way, the expected measured linear polarization when observing unpolarized sources of the same geometry was estimated. It turns out that the simulated linear polarizations are typically three times higher than the value determined from the beam maps for unresolved and unpolarized sources, only containing instrumental effects. In fact, it is of the same order of magnitude as are the actually observed polarizations. This clearly shows that the role of extended emission in the surroundings of the sources has to be taken into account in this kind of single-dish polarimetric observations. The effect of extended emission in polarized sidelobes on the derived source polarization has not been discussed in previous single-dish spectropolarimetry results (see Section~\ref{intro}). A comparison of those single-dish observations of the GK effect that have been published to date is further complicated by the fact that there is no overlap in sources and transitions.

Tentative evidence for intrinsically polarized line emission was found for G34.3+0.2 in the CS(2-1) transition, with $p_{\rm L}\lesssim1.5\%$. Here, the observed polarization in the blue-shifted line wing is clearly higher than the simulated instrumental response of an unpolarized source with the same geometry.

Future single-dish observations of weak linear polarization in molecular transitions have to take into account the role of extended emission and ideally would rely on even better receiver alignment and receivers which are as similar as possible. Our method of simulating the instrumental response for real source geometries enables polarimetric observations of sources with surrounding extended emission. However, the effect of extended emission complicates the search for polarized emission at the edge of molecular clouds where, as noted above, the polarization may be more easily observable because of more anisotropic optical depth.

\begin{acknowledgements} 
Based on observations carried out with the IRAM 30m telescope. IRAM is supported by INSU/CNRS (France), MPG (Germany) and IGN (Spain).
\end{acknowledgements}

\bibliographystyle{aa} 
\bibliography{bibmaster} 

\end{document}